\documentclass[aps,prl,twocolumn,superscriptaddress]{revtex4-1}
\usepackage{amsmath}
\usepackage{amssymb}
\usepackage{graphics}
\usepackage{graphicx}
\usepackage{float}
\usepackage{subfigure}
\usepackage{color}
\usepackage{hyperref}

\begin{document}
\title{Measuring the Boundary Gapless State and Criticality via Disorder Operator}

\author{Zenan Liu}
\affiliation{Guangdong Provincial Key Laboratory of Magnetoelectric Physics and Devices, State Key Laboratory of Optoelectronic Materials and Technologies, Center for Neutron Science and Technology, School of Physics, Sun Yat-Sen University, Guangzhou, 510275, China}

\author{Rui-Zhen Huang}
\affiliation{Department of Physics and Astronomy, Ghent University, Krijgslaan 281, S9, B-9000 Ghent, Belgium}

\author{Yan-Cheng Wang}
\email{ycwangphys@buaa.edu.cn}
\affiliation{Hangzhou International Innovation Institute, Beihang University, Hangzhou 311115, China}
\affiliation{Tianmushan Laboratory, Hangzhou 311115, China}

\author{Zheng Yan}
\email{zhengyan@westlake.edu.cn}
\affiliation{Department of Physics, School of Science and Research Center for Industries of the Future, Westlake University, Hangzhou 310030,  China}
\affiliation{Institute of Natural Sciences, Westlake Institute for Advanced Study, Hangzhou 310024, China}

\author{Dao-Xin Yao}
\email{yaodaox@mail.sysu.edu.cn}
\affiliation{Guangdong Provincial Key Laboratory of Magnetoelectric Physics and Devices, State Key Laboratory of Optoelectronic Materials and Technologies, Center for Neutron Science and Technology, School of Physics, Sun Yat-Sen University, Guangzhou, 510275, China}
\affiliation{International Quantum Academy, Shenzhen 518048, China}

\begin{abstract}
The disorder operator is often designed to reveal the conformal field theory  information in quantum many-body systems. By using large-scale quantum Monte Carlo simulation, we study the scaling behavior of disorder operators on the boundary in the two-dimensional Heisenberg model on the square-octagon lattice with gapless topological edge state. In the Affleck-Kennedy-Lieb-Tasaki  phase, the disorder operator is shown to hold the perimeter scaling with a logarithmic term associated with the Luttinger liquid parameter $K$. This effective Luttinger liquid parameter $K$ reflects the low-energy physics and CFT for (1+1)D boundary. At bulk critical point, the effective $K$ is suppressed but keeps finite value, indicating the coupling between the gapless edge state and bulk fluctuation. The logarithmic term numerically captures this coupling picture, which reveals the (1+1)D SU$(2)_1$ CFT and (2+1)D $O(3)$ CFT at boundary criticality. Our Letter paves a new way to study the exotic boundary state and boundary criticality.
\end{abstract}
\date{\today}
\maketitle

\textit{\color{blue}Introduction.-}
Quantum critical behaviors are important and long-historical topics in quantum many-body physics. The unconventional phase transitions can go beyond the Landau-Ginzburg-Wilson paradigm and have attracted many analytical and numerical studies, such as deconfined quantum critical point (DQCP)~\cite{Anders2007, Senthil2004,zhao2022scaling} and topological phase transition~\cite{Essin2014,JWMei2015,YCWang2017QSL,YCWang2018,ZY2020}. In addition, when the bulk undergoes the phase transition, the boundary can also show exotic critical behaviors, dubbed as surface critical behaviors (SCBs)~\cite{Zhang2017,ding2018,zhu2022exotic,JianPing2021}. The exotic surface criticality draws renewed attention which is induced by the coupling between the gapless boundary state and critical bulk fluctuation. The edge-bulk coupling is widely found to play an important role in the nontrivial surface criticality of different quantum antiferromagnetic Heisenberg models~\cite{Xu2022}. The gapless edge state composed of dangling spin-1/2 is considered as the origin of unconventional SCBs ~\cite{Zhang2017,ding2018,weber2018nonordinary,ding2023special,wang2023extraordinary}. However, the exotic boundary criticality is also observed without gapless edge state in the spin-1 model~\cite{weber2019}, which leads to more controversial problems in the boundary criticality and requires more useful detecting methods. 

In recent years, the nonlocal operators have been widely used to reveal the entanglement and categorical symmetry for quantum many-body systems, such as symmetry domain walls or field lines of emergent gauge fields~\cite{Wen2020,Liang2020,NUSSINOV2009977,yan2022triangular,yan2023emergent}. They pave a new path to probe the phases and phase transitions from the viewpoint of high-form symmetry or domain wall. The disorder operator is a nonlocal observable that is proposed to extract the high-form symmetry of quantum many-body systems~\cite{Wu2021,lake2018higherform,fradkin2017disorder}. It has been successfully used to detect the high-form symmetry breaking at Ising transition~\cite{zhao2021}. The current central charge can be extracted from the disorder operator at (2+1)D $O(2)$ and $O(3)$ phase transition in the CFT~\cite{Wang2021,Wang2022}. Fermion disorder operators are also designed to explore the universal feature of Fermi liquid, Luttinger liquid, fermion QCP and reflect the non-unitary conformal field theory (CFT)  of fermion DQCP~\cite{jiang2023versus,LiuF2023, liu2023disorder}. The disorder operator satisfies the new universal scaling behaviors, where the subleading logarithmic term reflects the general feature of CFT at the conformal invariant QCP. 

Since the exotic boundary criticality is combined by both edge and bulk modes which seems to contain either the CFT information of edge excitation or bulk branch, how to extract the composite CFTs at the special criticality is a core problem to be solved when we want to study the mechanism of the exotic edge class. 
In this Letter, we first introduce the disorder operator to probe the boundary state and boundary criticality by taking the $S=1/2$ Heisenberg model on the square-octagon lattice as an example.  The scaling behaviors also hold the perimeter law with subleading logarithmic term in the Affleck-Kennedy-Lieb-Tasaki (AKLT) phase. This logarithmic term is related to the Luttinger liquid (LL) 
parameter $K$ at small angle value, which demonstrates that the boundary is governed by LL in the IR limit. The most intriguing case is that the logarithmic term $s(\theta)$ can capture the LL parameter $K$ for boundary and current central charge $J$ for bulk at QCP. This logarithmic term scaling is different from the bulk disorder operator at (2+1)D QCP, which provides a new tool to understand the boundary physics.  
\begin{figure}[t!]
\centering
\includegraphics[width=0.48\textwidth]{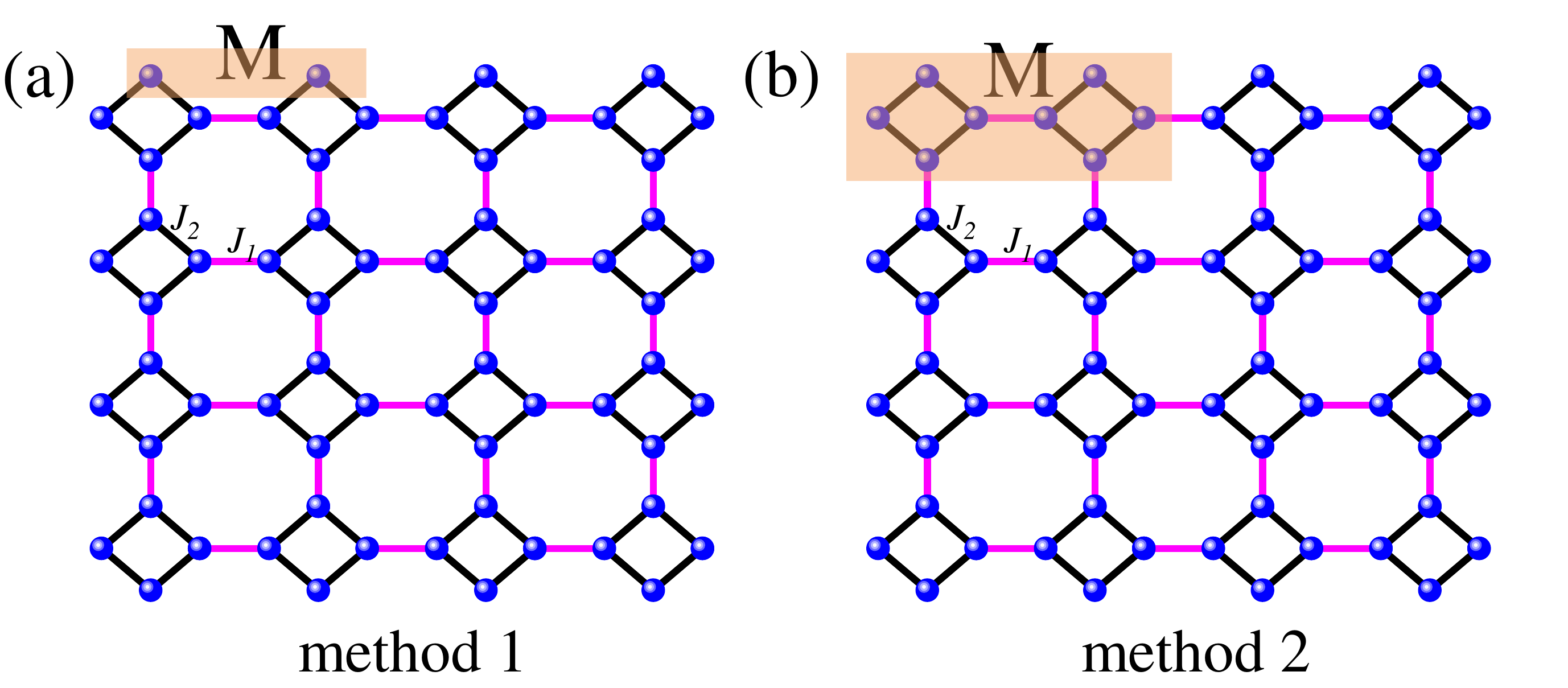}
\vspace{5pt}
\includegraphics[width=0.48\textwidth]{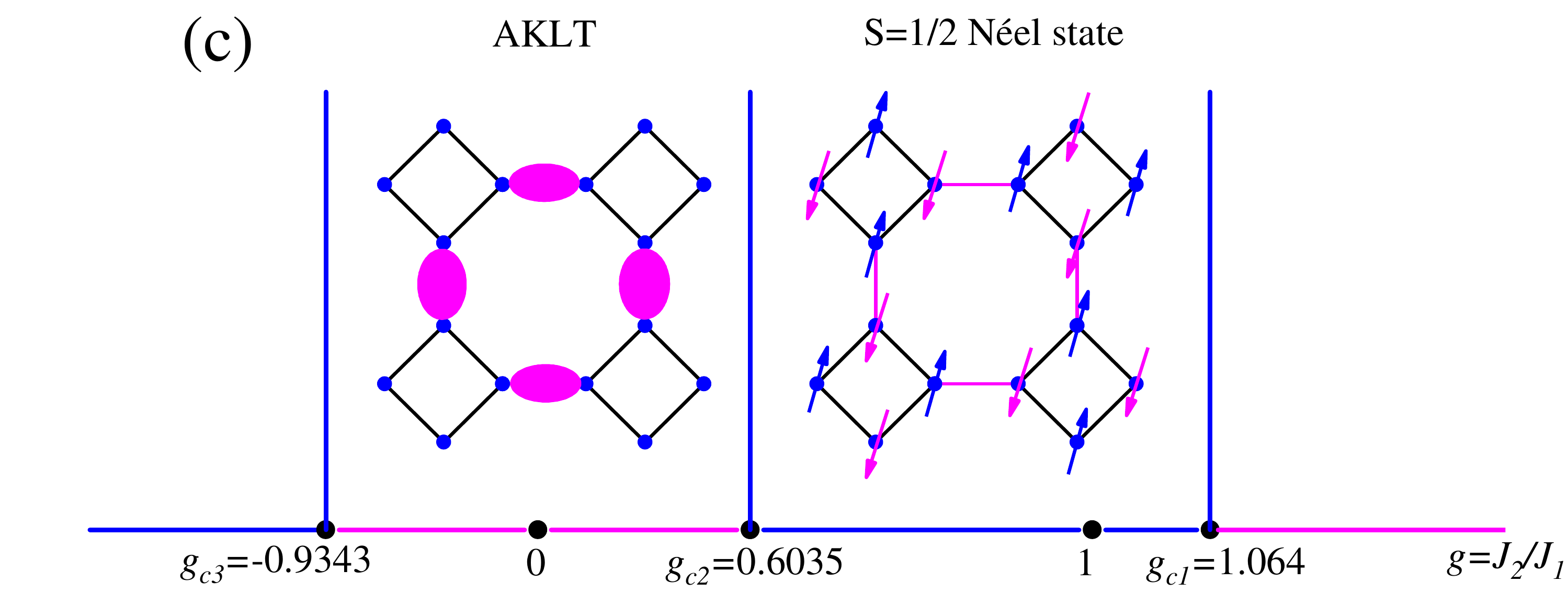}
\caption{ $J_1$-$J_2$ Heisenberg model on the square-octagon lattice. There are two different measurement region $M$s (orange region) for disorder operator on the edge, dubbed as method 1 (a) and method 2 (b). (c) Ground-state phase diagram of this model. We mainly consider the phase transition between the AKLT state and $S=1/2$ N\'eel state}
\label{fig-lattice}
\end{figure}

\textit{\color{blue}Model and method.-}
We investigate the spin-$\frac{1}{2}$ $J_1-J_2$ Heisenberg model on a square-octagon lattice via quantum Monte Carlo (QMC) simulations~\cite{Sandvik1999,Sandvik2010,sandvik2019}, also known as the AKLT model.
\begin{equation}
\begin{split}
H = J_{1}\sum_{\langle ij\rangle}\mathbf{S}_{i}\cdot \mathbf{S}_{j}+J_{2}\sum_{\langle ij\rangle'}\mathbf{S}_{i}\cdot \mathbf{S}_{j}
\end{split}
\end{equation}
where $J_1$ is the inter-unit-cell coupling and $J_2$ is the intra-unit-cell coupling. We define $g=J_2/J_1$ and set $J_1=1$. This model can host rich phase diagram via tuning the coupling $J_2$,  which includes the $S=2$ N\'eel state, AKLT state, $S=1/2$ N\'eel state, and plaquette valence bond crystal ~\cite{Liu2022, Zhang2017,Huang2022}. These phases are all separated by $O(3)$ quantum critical points. The AKLT state is a symmetry-protected topological phase whose boundary is protected by translation symmetry and spin-rotation symmetry~\cite{aklt1987,Gu2009}. So the gapless boundary is governed by an effective $S=1/2$ Heisenberg chain in the low-energy physics. At the bulk $O(3)$ quantum critical point, the boundary gapless mode is coupled to the bulk fluctuation which induces the unconventional boundary criticality. This demonstrates that the boundary can have richer physics and critical behaviors than bulk. 

\textit{\color{blue} Disorder operator.-} For a quantum system with U(1) symmetry, the disorder operator can be constructed by U(1) rotation angle $U(\theta)=\prod_i e^{i\theta(S^{z}_{i}-\frac{1}{2})}$, where $S^{z}_{i}$ is the U(1) charge on site $i$. Given a region $M$ on the lattice, we can define the disorder operator $X_M(\theta)=\prod_M e^{i\theta(S^{z}_{i}-\frac{1}{2})}$. The ground state expectation
\textit{\(|\langle X(\theta) \rangle|\)}  is the module of  $\langle X(\theta) \rangle$ defined as the disorder parameter that can extract the order and high-form symmetry of the disorder phase. The scaling behaviors of $X(\theta)$ rely on whether the phase is ordered or disordered. In the disordered phase, $|\langle X(\theta) \rangle|$ is proportional to $e^{-a(\theta)l}$ ~\cite{Wang2021,Wu2021,Estienne_2022}, where $l$ is the perimeter of the region $M$, meaning it obeys the perimeter law.  In the U(1) symmetry breaking ordered phase, it usually satisfies $|\langle X(\theta) \rangle| \sim e^{-b(\theta)l\ln l}$ ~\cite{Wang2021,Wu2021,lake2018higherform}. More interestingly, the logarithmic correction term will appear in the scaling behavior of $|\langle X(\theta) \rangle|$ at QCP. The previous analytical and numerical works pointed out that $|\langle X(\theta) \rangle|$ can hold the following form for a rectangle region at (2+1)D QCP~\cite{Wang2021,Wang2022,Wu2021}:
\begin{equation}
\begin{split}
\label{eq1}
\ln|\langle X(\theta) \rangle|=-a_1 l+s\ln l+a_0
\end{split}
\end{equation}
where all the coefficients are as functions of $\theta$. This logarithmic term $s(\theta)$ originates from the corner of the region $M$, which is also a universal function of operator angle $\theta$
and the open angle $\alpha$ of the corner in region $M$ ($\alpha=\pi/2$ in rectangle region). This corner correction $s$ can be used to detect the universal feature at QCP. Some previous works~\cite{Wang2021,Wang2022} have suggested that this term has a simple form as $s(\theta)=\frac{C_J}{(4\pi)^2}\theta^2$ ($\theta \rightarrow{0}$), where $C_J$ is the current central charge depending on the universality in the CFT. For a pure (1+1)D gapless system, the leading term of $|\langle X(\theta) \rangle|$ becomes a logarithmic term rather than a linear term $l$, due to the fact that the boundary of the disorder operator is a pointlike domain wall. So the scaling of $|\langle X(\theta) \rangle|$ takes the following form~\cite{jiang2023versus, giamarchi2003},

\begin{equation}
\begin{split}
\label{eq2}
\ln|\langle X(\theta) \rangle|=s\ln l+a_0
\end{split}
\end{equation}
Here, the logarithmic term can be connected to the LL parameter $K$ in the region of small angle value, which can be derived from the LL theory~\cite{giamarchi2003}.
In the LL system, such as spin-1/2 $XXZ$ chain, the analytical results have suggested $s(\theta)=-\frac{K}{2\pi^2}\theta^2$ ($\theta \rightarrow{0}$), where $K$ is the LL parameter of systems. For a one-dimensional spin-1/2 $XXZ$ chain, $K$ can be exactly solved from the anisotropic parameter $K=(\pi/2)/[\pi-arccos(\Delta)]$ ($\Delta$ is the anisotropic parameter)~\cite{Yang2007}. In this way, the disorder operator provides a simple way to measure the LL parameter.

In the past, the disorder operator is usually defined on the bulk, which captures the universal feature of bulk criticality or the bulk phase. Meanwhile, it has been found that the boundary contains richer physics especially at QCP~\cite{ding2018,deng2005surface,zhu2021surface,weber2018nonordinary,Toldin2021,Max2022}, but it has not been explored widely via non-local operators. When the disorder operator is defined on the boundary, similarly, it seemingly can also extract the universal feature of the gapless state and even boundary criticality, which should be demonstrated through a logarithmic term.  For  disorder operator on edge, it has a perimeter $l$ of the measurement region $M$ between the edge and bulk, which contributes to the leading term $\sim l$ in the disorder phase~\footnote{For a pure (1+1)D system, it has no perimeter term because there is no coupling between the (1+1)d system and an external bulk.}. The perimeter $l$ of region $M$ gives contribution to the leading term for the disorder operator. So it may take a similar form as Eq.~(\ref{eq2}). The logarithmic term actually captures the boundary physics.  At the same time, two kinds of definitions for the disorder operator on the edge are performed in Figs.~\ref{fig-lattice} (a) and ~\ref{fig-lattice} (b). We use method 1 to measure the boundary disorder operator in most of the QMC simulations, as it can reflect the effective Heisenberg chain on the boundary in our previous study~\cite{Liu2022}  

\begin{figure}[t!]
\centering
\includegraphics[width=0.45\textwidth]{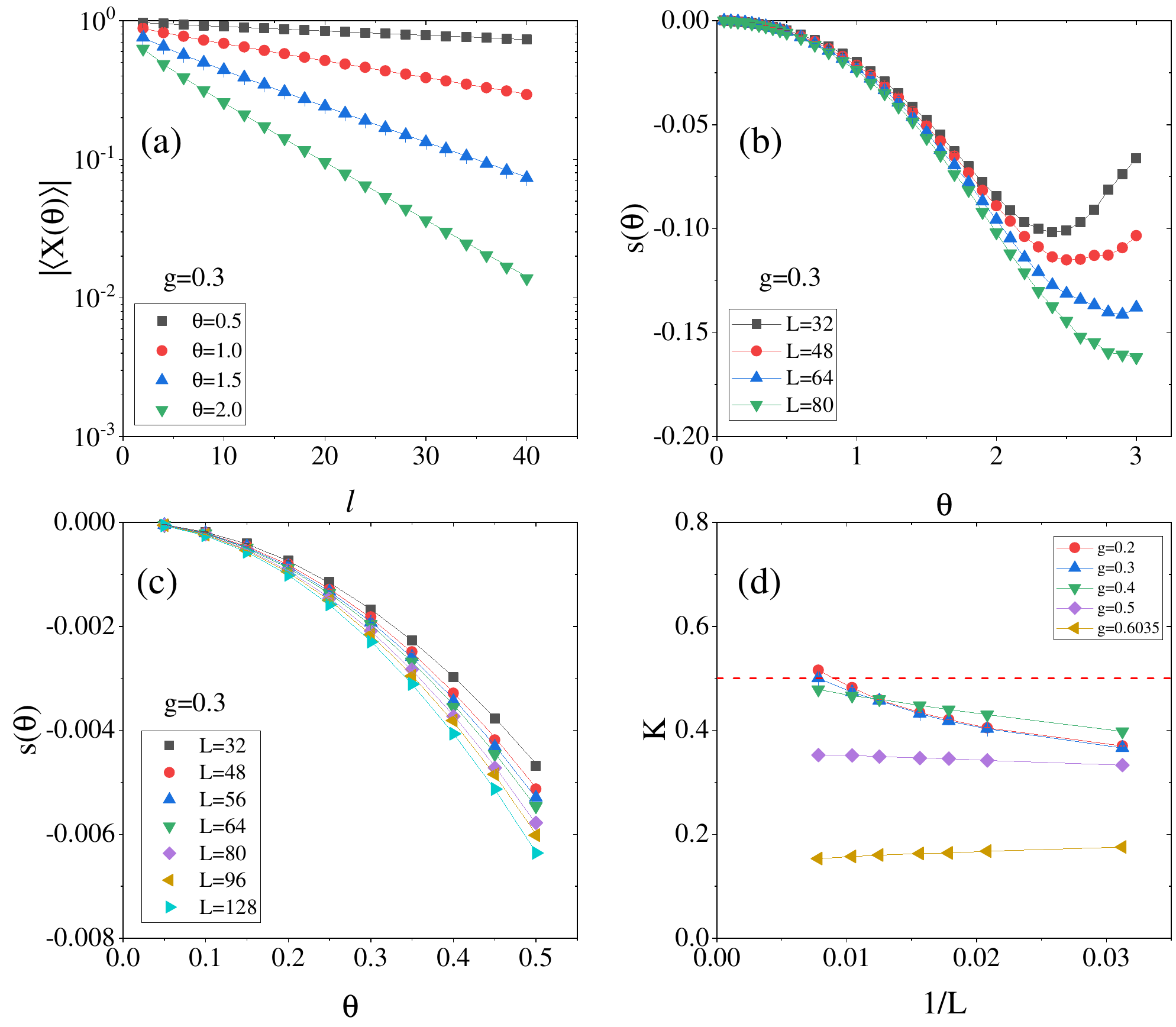}
\caption{(a) Disorder parameter $|\langle X(\theta) \rangle|$ as a function of edge length $l$ with system size $L=80$ for $g=0.3(a)$. (b) shows the subleading term s($\theta$) obtained from (a), with system size $L=32, 48, 64, 80$. (c) The coefficient of the logarithmic correction $s(\theta)$ for small value of $\theta$ with system size $L=32, 48, 64, 80, 96, 128$ at $g=0.3$. (d) The finite size extrapolation of Luttinger liquid parameter $K$ with increasing system size $L$ with different $g$.}
\label{Fig.s1}
\end{figure}

\textit{\color{blue}Results.-}
In the 1D $XXZ$ chain, the disorder operator successfully extracts the LL parameter $K$, which is well consistent with the theoretical value (more details can be found in Appendix A).  For the 2D AKLT model, the gapless boundary is equivalent to an effective 1D Heisenberg chain in the AKLT phase, which can be considered as a (1+1)D system due to the gapped bulk. Naturally, the disorder operator on the boundary may have similar scaling behavior as it in pure (1+1)D systems.  As shown in Fig.~\ref{Fig.s1}, we obtain the boundary disorder operator value with $g=0.3$ and $L=80$. The scaling behavior satisfies the Eq.~(\ref{eq1}), where the leading term $l$ originates from the touch between the boundary disorder operator and bulk in the disorder phase. 
The small angle value of $s(\theta)$ can be fitted well by $s(\theta)=-\frac{K}{2\pi^2}\theta^2$ rather than $s(\theta)=\frac{C_J}{(4\pi)^2}\theta^2$ at small angle value (Fig.~\ref{Fig.s1}), which is different from the (2+1)D disorder operator result. The LL parameter $K$ is obtained from the fitting $s(\theta)$ with system size from $L=32$ to $128$ with $\beta=2L$ ($\beta=4L$ for $g=0.2$ due to the small energy scale of interaction). For small $g$, the extrapolation of fitting $K$ will converge to 0.5, which is consistent with the LL parameter of $S=1/2$ Heisenberg chain. Thus, the unique (1+1)D SU$(2)_1$ can be well captured in the deep AKLT state. When $g$ gets closed to QCP, the extrapolation of $K$ becomes smaller, such as $g=0.5$ and $g=0.6035$ results in Fig.~\ref{Fig.s1}(d). The feature of LL on the boundary becomes weaker and weaker, and the boundary cannot be considered as a pure (1+1)D system which is not described well by the unique (1+1)D CFT anymore.

\begin{figure}[t!]
\centering
\includegraphics[width=0.45\textwidth]{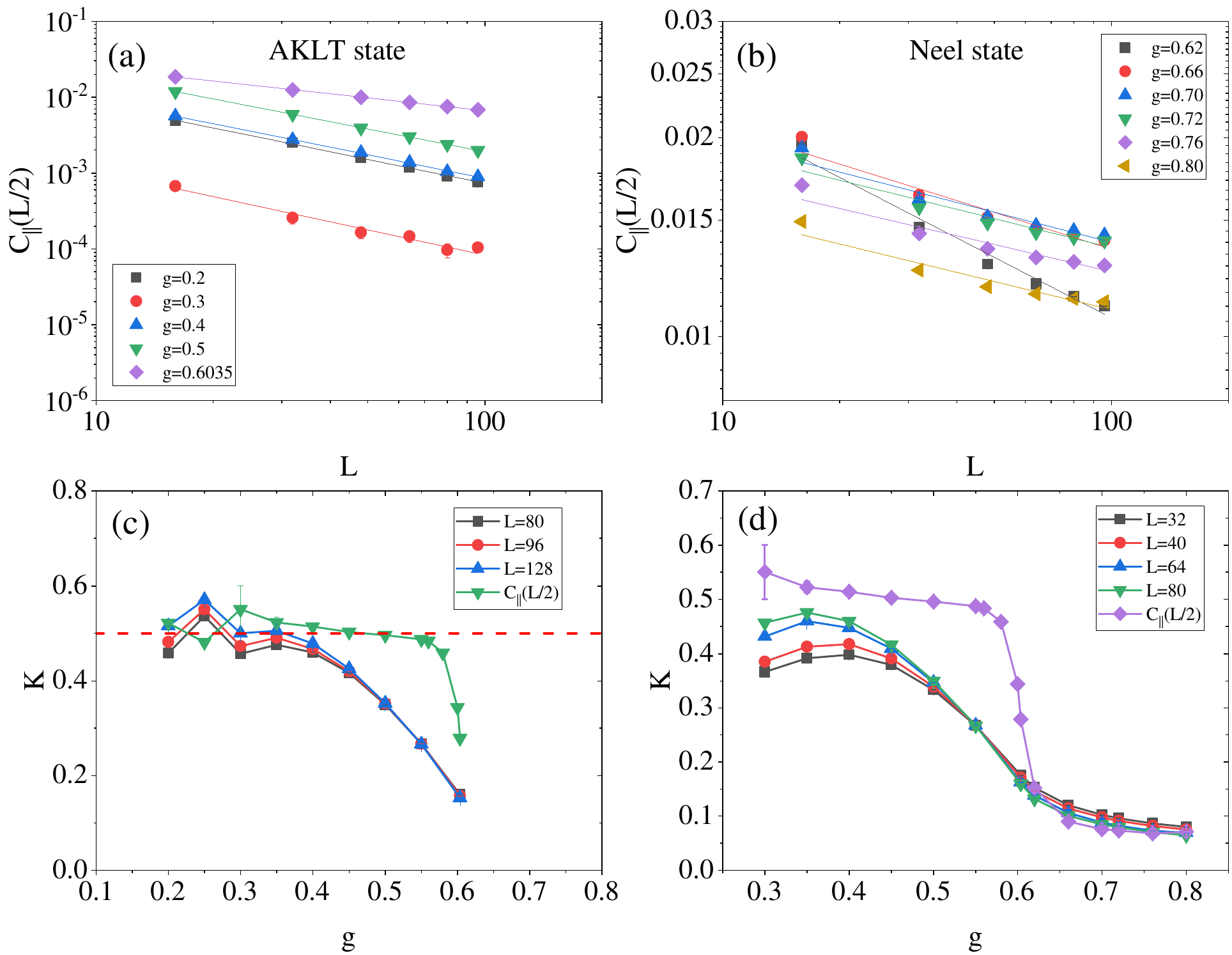}
\caption{Spin correlation functions $C_{||}(r)$ at $r=L/2$ on the surface are shown at the (a) AKLT state and (b) N\'eel state. (c) The Luttinger parameter $K$ from disorder operator (method 1) in the AKLT state from $L=80$ to 128. (d) The change of $K$ from disorder operator (method 1) at quantum phase transition from AKLT state to N\'eel state from $L=32$ to 80}.
\label{Fig.s3}
\end{figure}

\begin{figure*}[htbp]
\centering
\includegraphics[width=0.96\textwidth]{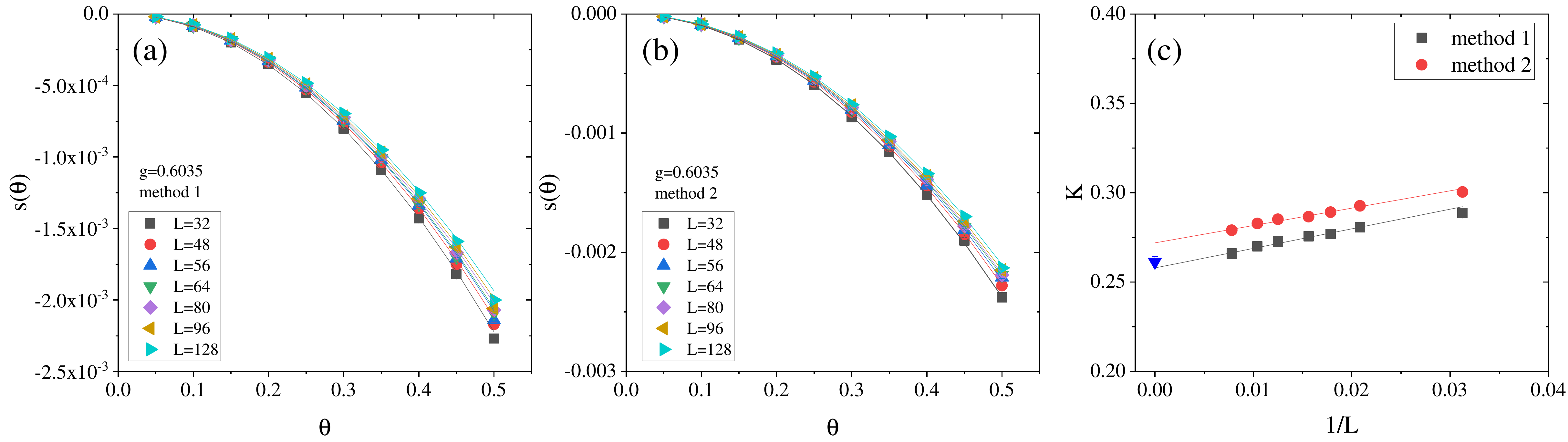}
\caption{The refitting logarithmic correction $s(\theta)$ for small value $\theta$  for (a) method 1 and (b) method 2 at quantum critical point with system size from 32 to 128. (c) The refitting LL parameter obtained from disorder operator are compared with the correlation function. The blue dot is the result of correlation function.}
\label{Fig.s4}
\end{figure*}

In order to verify the results of the disorder operator, we calculate the boundary spin correlation function $C_{||}(r)$ to obtain the effective LL parameter. 
Here $C_{||}(r)$ means that the spins we measure are along the boundary with $r$ parallel to the boundary.
According to the LL theory, the two-point spin correlation function in $S=1/2$ Heisenberg chain satisfies $C(L/2) \sim L^{-2K}$~\cite{giamarchi2003}($L$ is the system size and $K$ is the LL parameter), as depicted in Fig.~\ref{Fig.s3} (a) and \ref{Fig.s3} (b). Although the boundary for N\'eel state can not be regarded as a pure (1+1)D system, the effective $K$ can still be obtained from the correlation function. In the AKLT state, the fitting $K$ obtained from the disorder operator is consistent with the correlation function in small $g$, as Fig.~\ref{Fig.s3}(c) illustrated for method 1 of the disorder operator (system size up to $L=128$). As $g$ gets closed to QCP, $K$ obtained from the correlation function is close to 0.5 and suddenly decreases until around QCP. However, The $K$ fitted by the disorder operator gradually decreases and deviates from 0.5 when $g>0.4$, suggesting that the bulk fluctuation influences the disorder operator near the QCP.  
The effective $K$ from the correlation function looks sharp while the disorder operator looks much smoother.  
As Fig.~\ref{Fig.s3}(d) shows for method 1 of the disorder operator (system size up to $L=80$), from the AKLT state to the N\'eel state, the fitting $K$ gradually becomes smaller and decays to 0 as g increases, indicating the boundary cannot be well described by (1+1)D LL theory due to the coupling between the edge and bulk.
Meanwhile, the disorder operator clearly shows that the boundary goes from the (1+1)D physics to the (2+1)D physics, indicating the bulk displays a phase transition.

Moreover, the unconventional boundary criticality behavior is induced when the gapless topological edge mode is coupled to the gapless bulk fluctuation~\cite{Xu2022}. 
When we revisit the disorder operator at QCP, we find that the bulk fluctuation has a significant influence to the disorder operator physically. 
At bulk QCP, the logarithmic correction $s(\theta)$ in the scaling of disorder operator captures the QCP feature, which is contributed by the corners of measurement region $M$. For the boundary disorder operator, we note that there are also two corners in the region $M$ with opening angle $\pi/2$. Because the bulk QCP belongs to the $O(3)$ universality class, these corner contributions can be equal to one half of $\frac{C_J}{(4\pi)^2}$ at small angle value in CFT. Therefore, according to this boundary criticality feature, we suggest that the logarithmic term $s$ takes the following form at small angle value:
\begin{equation}
\label{eq3}
s(\theta)=\Big( -\frac{K}{2\pi^2}+\frac{C_J}{2(4\pi)^2}\Big)\theta^2
\end{equation}

Here the LL parameter $K$ captures the (1+1)D SU$(2)_1$ CFT and $C_J$ captures the (2+1)D $O(3)$ CFT. As we know, the theoretical value of $\frac{C_J}{(4\pi)^2}$ is 0.01147 from numerical bootstrap~\cite{Poland2019} in the $O(3)$ CFT. When we refit the logarithmic term $s(\theta)$ via Eq.~(\ref{eq3}) with the theoretical value of $\frac{C_J}{(4\pi)^2}$, it is surprising to find that the disorder operator of fitting $K$ is almost consistent with the results of the correlation function as shown in Fig.~\ref{Fig.s4}. The extrapolation of fitting $K$ is 0.2579(2), which is very close to 0.261(3) obtained from $C(L/2)$. Also, we use measurement method 2 to capture the effective $K$ at QCP (Fig.~\ref{Fig.s4}). The fitting $K$ from method 2 converges to 0.2720(2) as $L \rightarrow \infty$, which also agrees well with the results of $C(L/2)$. Because the boundary couples with the bulk, there may be high-order corrections in the subleading term. Therefore, it is reasonable that the fitting $K$ from the disorder operator may deviate a little from the correlation function.

\textit{\color{blue}Discussion.-}
The boundary criticality can be controlled by the bulk interaction or boundary interaction, which is also well captured by the boundary conformal field theory ~\cite{CARDY1984,CARDY1986,CARDY1989}. 
The boundary of the AKLT state is governed by the unique (1+1)D SU$(2)_1$ CFT, which exactly contributes to the logarithmic term. 
As bulk approaches the critical point, The SU$(2)_1$ CFT is unstable against the bulk fluctuation, where LL parameter $K$ is suppressed by the N\'eel order. This boundary criticality is a special transition fixed point under the renormalization group, which is similar to the phase transition between N\'eel order and valence bond solid order in a one-dimensional system with non-local interaction~\cite{Xu2022,Anders2010,yang2020deconfined}. 
However, it is difficult to derive the analytical expression for $s(\theta)$ from the boundary critical correlation function at QCP. According to the physics picture for the disorder operator, we suggest that it will take the above form as Eq.~(\ref{eq3}). The correlation function results further confirm this conjecture that it captures the composite CFT at QCP as Fig.~\ref{Fig.s4} shows. As we know, the Luttinger parameter of the edge mode can also be extracted from the correlation function, meanwhile the Luttinger parameter K represents the (1+1)D SU$(2)_1$ CFT information. According to the numerical results, the Luttinger parameter $K$ from disorder operator is consistent with the value of correlation function. This means that the disorder operator indeed captures the physics and CFT information of (1+1)D SU$(2)_1$ boundary~\cite{jiang2023versus} (details about the relationship between disorder operator and Luttinger parameter $K$ can be found in the Appendix of Ref.~\cite{jiang2023versus}). For (2+1)D Heisenberg model, the boundary gapless state often exists at QCP that couples to the bulk mode, which leads to the unconventional SCB. Therefore, the LL parameter $K$ keeps finite value in the thermodynamic limit. Moreover, if the gapless edge state is suppressed to 0 by bulk fluctuation at QCP, the LL parameter $K$ will also decay to 0 rather than the finite value at QCP. It is interesting to explore how LL parameter $K$ disappears while the central charge term $\frac{C_J}{(4\pi)^2}$ remains at QCP in this case, which we leave for future work.  

\textit{\color{blue}Conclusion.-}
In summary, we first apply the disorder operator to study the boundary state and boundary criticality in the two-dimensional AKLT model. 
In the AKLT state, the boundary gapless state can be considered as an effective $S=1/2$ Heisenberg chain, which has a well-defined LL parameter $K$.  The effective $K$ can be obtained from the logarithmic term $s(\theta)$ at small angle value, which is close to the $K$ obtained from the correlation function. As the bulk undergoes the phase transition between AKLT state and N\'eel state, the LL parameter $K$ gradually decreases and then decays to about 0 in the N\'eel state, which is clearly shown in the disorder operator and correlation function. More importantly, the disorder operator can capture the current central charge $C_J$ of $O(3)$ CFT and LL parameter $K$ of (1+1)D SU$(2)_1$ CFT, which can be confirmed via correlation function too. These results demonstrate that the boundary gapless state couples to the bulk fluctuation in the AKLT model, which can be numerically considered as composite CFT in the disorder operator. Our Letter shows that the non-local disorder operator can detect the boundary state and criticality, which provides a new window to understand the CFT information on the boundary.

\textit{\color{blue}Acknowledgements.-}
We wish to thank Zi Yang Meng and Meng Cheng for useful discussions. Z.L. and D.X.Y. are supported by NKRDPC-2022YFA1402802, NKRDPC-2018YFA0306001, NSFC-92165204, NSFC-11974432, Leading Talent Program of Guangdong Special Projects (201626003), Guangdong Provincial Key Laboratory of Magnetoelectric Physics and Devices (No. 2022B1212010008), and Shenzhen Institute for Quantum Science and Engineering (No. SIQSE202102). Z.L. also thanks the academic visit at Westlake University. Z.Y. thanks the support from the start-up funding of Westlake University. Y.C.W. is supported by Zhejiang Provincial Natural Science Foundation of China (No. LZ23A040003) and the High-Performance Computing Centre of Hangzhou International Innovation Institute of Beihang University. R.Z.H. is supported by a postdoctoral fellowship from Ghent University - Special Research Fund (BOF). The authors also acknowledge Beijng PARATERA Tech Co.,Ltd., the HPC Center of Westlake University for providing HPC resources, and National Supercomputer Center in Guangzhou.


%

\clearpage
\appendix
\section{Appendix A: Disorder operator in $XXZ$ chain}
\begin{figure}[htbp]
\centering
\includegraphics[width=0.45\textwidth]{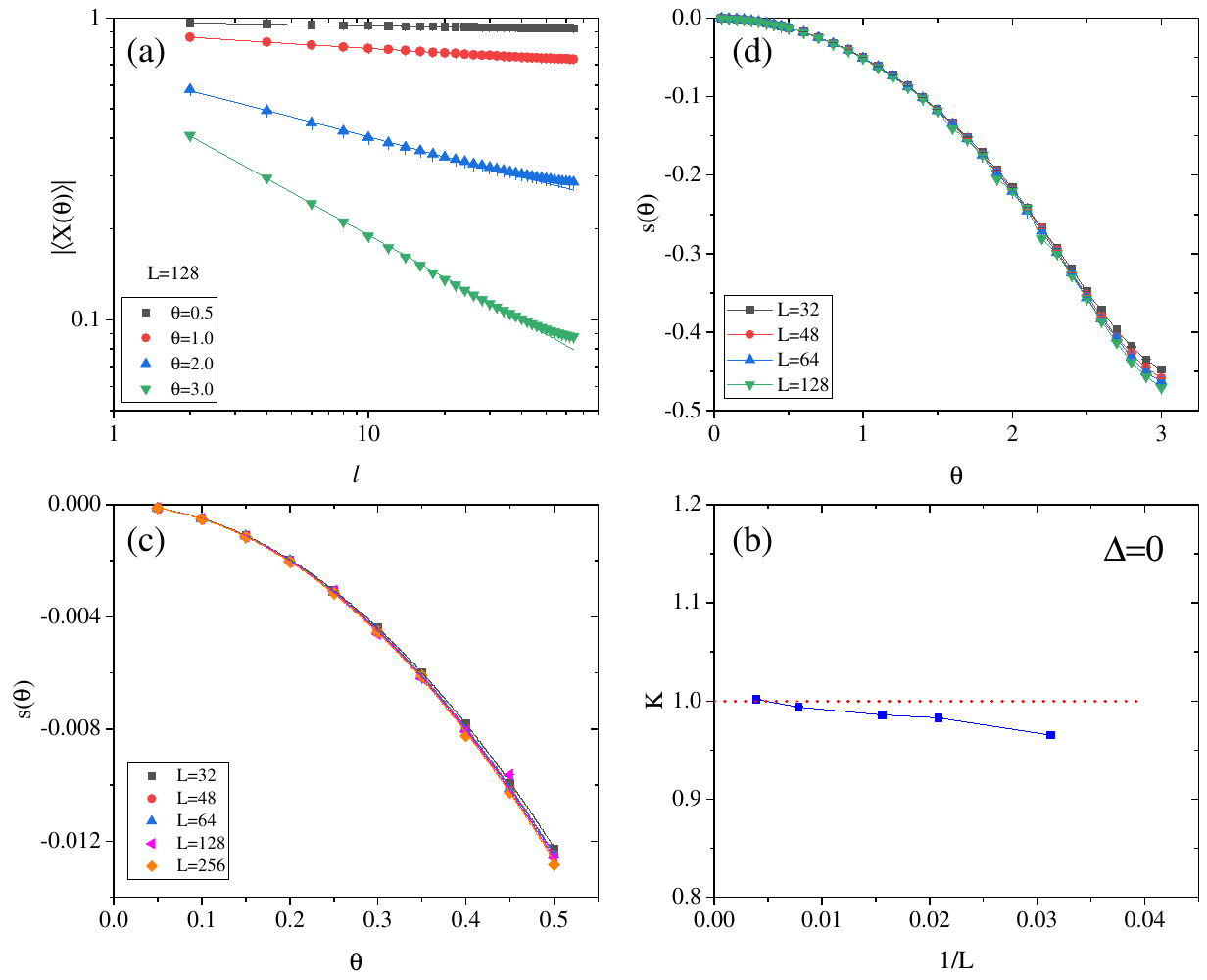}
\caption{Disorder parameter $|\langle X(\theta) \rangle|$ of $S=1/2$ $XXZ$ chain with $\Delta=0$. (a) $|\langle X(\theta) \rangle|$ as a function of system size $L$. (b) The logarithmic correction $s(\theta)$ obtained from (a). (c) Small angle value of $s(\theta)$. (d) Finite size extrapolation  of Luttinger liquid parameter $K$.}
\label{Fig.h1}
\end{figure}

In order to show the disorder operator can extract the feature of a (1+1)D system, we first perform the numerical calculations in the $S=1/2$ $XXZ$ chain with $\beta=2L$. Generally, the $XXZ$ chain can be mapped to the free fermion chain with interaction by a Jordan-Wigner transformation. The scaling of the disorder operator should be equivalent to the fermion disorder operator in the free fermion chain. In the $XXZ$ chain with $\Delta=0$, $|\langle X(\theta) \rangle|$ satisfies the scaling behavior of Eq.~(\ref{eq2}), as shown in Fig.~\ref{Fig.h1}.  And now the system becomes an $XY$ chain with $\Delta=0$, which can be mapped to a free fermion chain without interaction. Therefore the logarithmic term $s(\theta)$ can be obtained from the fitting of Eq.~(\ref{eq2}) at any angle value. As we know, $s(\theta)$ holds the simple form $s(\theta)=-\frac{K}{2\pi^2}\theta^2$ at small angle value~\cite{jiang2023versus}, where the LL parameter $K=1$ when $\Delta=0$, according to the analytical results. The fitting LL parameter $K$ is found to converge to 1, which is well consistent with the theory.

\begin{figure}[htbp]
\centering
\includegraphics[width=0.45\textwidth]{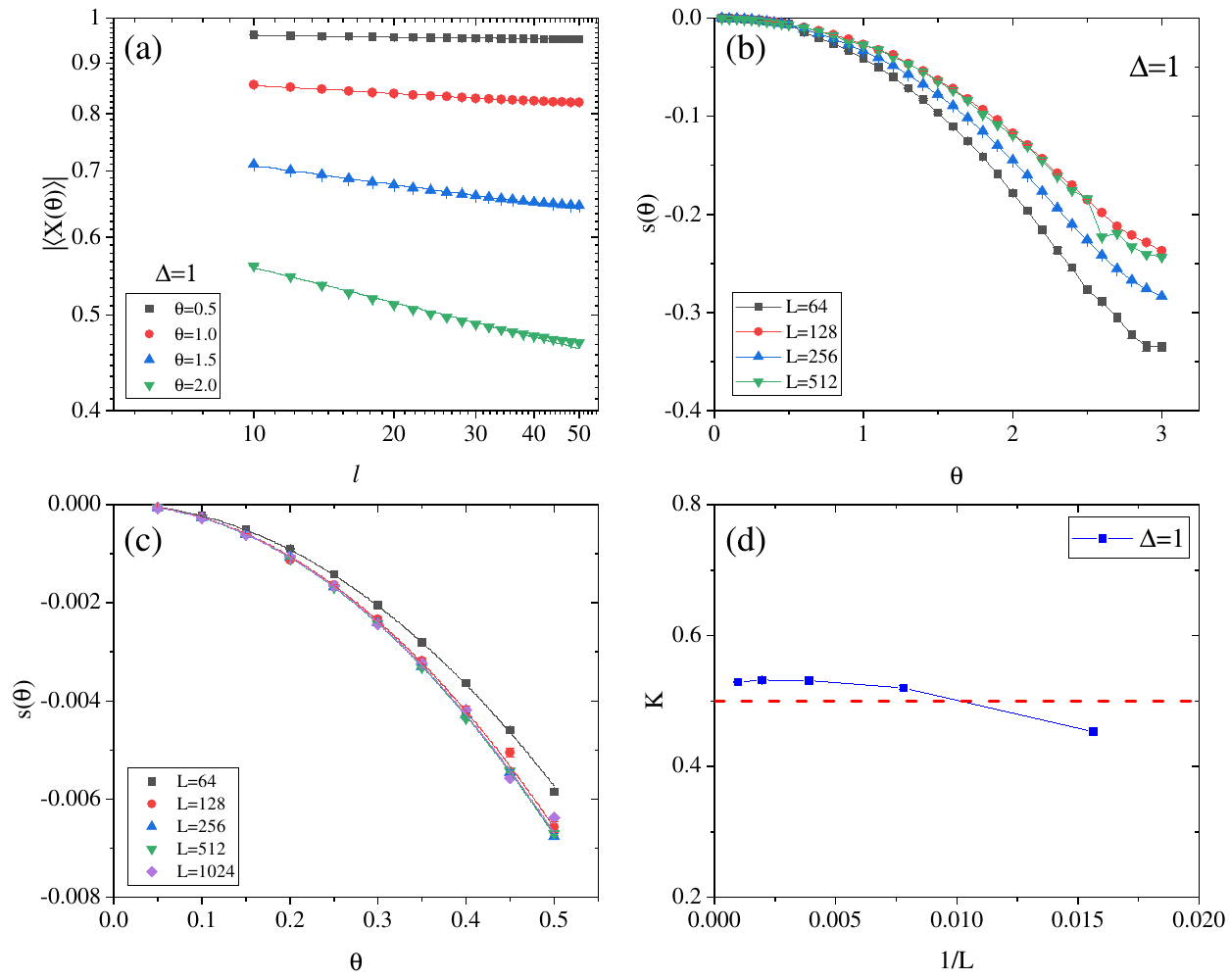}
\caption{Disorder parameter $|\langle X(\theta) \rangle|$ of $S=1/2$ $XXZ$ chain with $\Delta=1$. (a) $|\langle X(\theta) \rangle|$ as a function of system size $L$. (b) The logarithmic correction $s(\theta)$ obtained from (a). (c) Small angle value of $s(\theta)$. (d) Finite size extrapolation  of Luttinger liquid parameter $K$.}
\label{Fig.h2}
\end{figure}

 And the $XXZ$ chain becomes an isotropic Heisenberg chain which is also at a Kosterlitz-Thouless  phase transition with $\Delta=1$. The scaling of $|\langle X(\theta) \rangle|$ can be fitted by Eq.~(\ref{eq2}), in which the logarithmic term $s$ reflects the LL parameter. The fitting LL parameter $K$ converges to 0.52, which is close to the theoretical value $K=0.5$, as Fig.\ref{Fig.h2} shows. For the Heisenberg chain, there is a marginal operator in the field form of Hamiltonian~\cite{giamarchi2003}, which causes much strong finite size effect in the simulations. So the extrapolation of fitting $K$ is hard to converge to exact 0.5.

\section{Appendix B: Scaling behaviors in the AKLT state}

\begin{figure*}[htbp]
\centering
\includegraphics[width=0.9\textwidth]{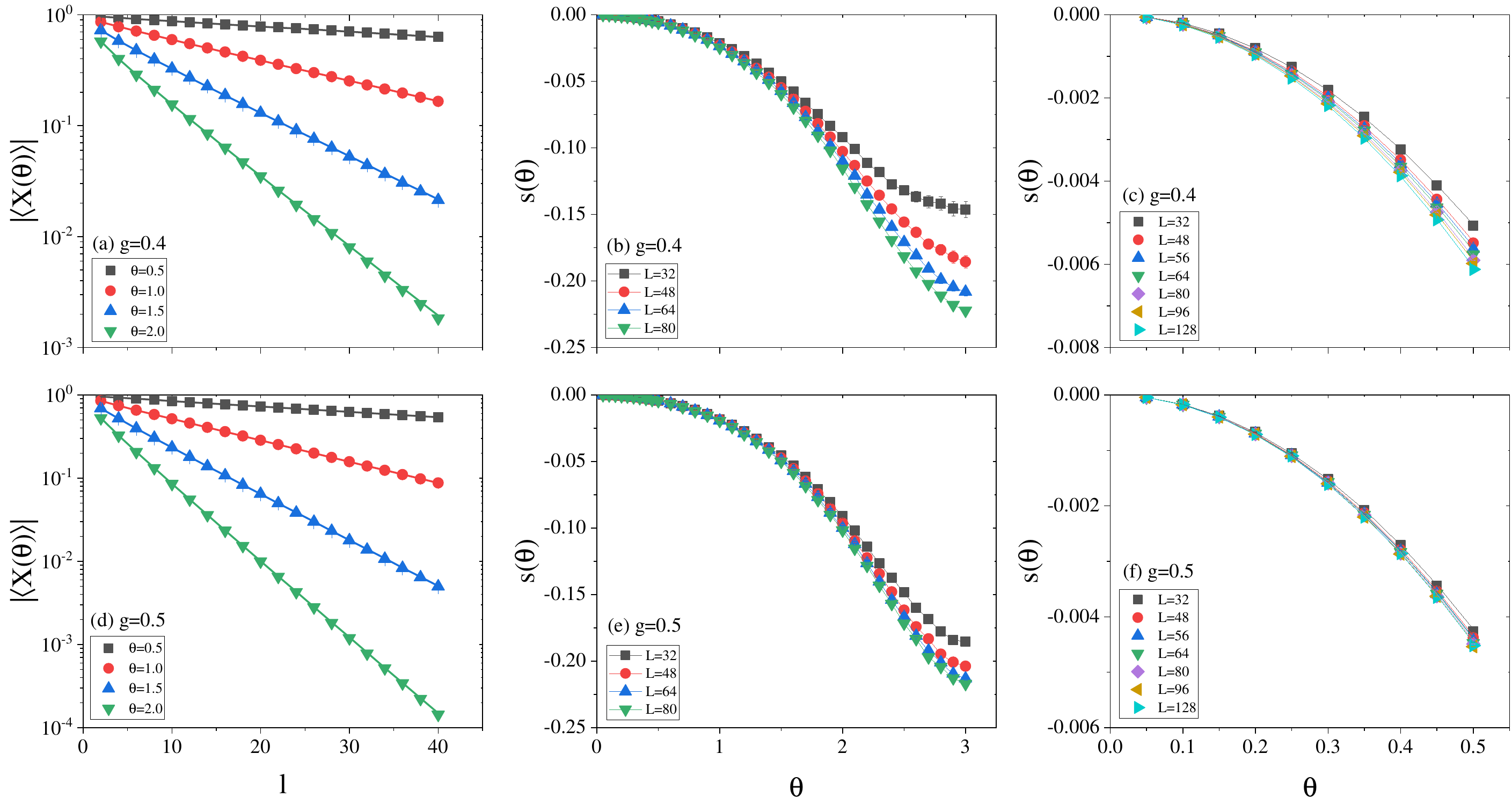}
\caption{ Disorder parameter $|\langle X(\theta) \rangle|$ as a function of edge length $l$ with system size $L=80$ for (a) $g=0.4$ and (d) 0.5.  The subleading term s($\theta$) obtained from the scaling of disorder operator, with system size $L=32, 48, 64, 80$ for (b) $g=0.4$ and (e) 0.5. The coefficient of the logarithmic correction $s(\theta)$ for small value of $\theta$ with system size $L=32, 48, 64, 80, 96, 128$ at $g=0.4$ (c) and $g=0.5$ (f). }
\label{Fig.s8}
\end{figure*}

For $g=0.4$ and $g=0.5$, the scaling behavior of the disorder operator also satisfies the same form as Eq.~(\ref{eq1}) as shown in Figs.~\ref{Fig.s8} (a) and \ref{Fig.s8} (c) with $\beta=2L$. This further demonstrates that the leading term is the perimeter $l$ in the AKLT phase. And the logarithmic term $s(\theta)$ is negative at full angle value $\theta$, which reflects the (1+1)D physics. Similar to the $g=0.3$ case, the logarithmic term can be used to extract the effective LL parameter $K$ by fitting the same equation at small angle limit [Figs.~\ref{Fig.s4} (c) and ~\ref{Fig.s4} (f)]. These results further confirm that the disorder operator can capture the only one (1+1)D SU$(2)_1$ CFT in the deep AKLT state, which is distinguished from the case of composite CFT at QCP.

\end{document}